\documentclass[11pt]{article}
\usepackage{amsfonts}
\usepackage{amsmath}
\usepackage{amssymb}
\usepackage{graphicx}
\usepackage{array}
\usepackage{xcolor}
\usepackage{booktabs}
\usepackage{multirow}
\usepackage{makecell}
\usepackage{float}
\usepackage{appendix}
\usepackage{cite}
\usepackage{subcaption} 
\pdfoutput=1 

\begin{document}
\begin{center}
\large \textbf{Interplay of Lyapunov exponents, phase transitions and chaos bound in nonlinear electrodynamics black hole}
\end{center}

\begin{center}
Chuanhong Gao$^{a*}$ \footnote{E-mail: {chuanhonggao@hotmail.com}},
Chuang Yang$^{b}$ \footnote{E-mail: {chuangyangyc@hotmail.com}},
Tetvui Chong$^{a}$ \footnote{E-mail: {chongtetvui@gmail.com }},
Deyou Chen$^{b*}$ \footnote{E-mail: {deyouchen@hotmail.com}} 

\vspace{0.5em}

\textsuperscript{a}\textit{Faculty of Engineering and Quantity Surveying, INTI International University, Persiaran Perdana BBN, Putra Nilai, Nilai 71800, Negeri Sembilan, Malaysia}

\textsuperscript{b}\textit{School of Science, Xihua University, Chengdu 610039, China}

\textsuperscript{*}\textit{Corresponding authors}
\end{center}

\abstract{In this paper, we investigate Lyapunov exponents of chaos for both massless and charged particles around a non-linear electrodynamics black hole, and explore their relationships with a phase transition and a chaos bound of this black hole. Our results indicate that these exponents can effectively reveal the phase transition. Specifically, during the phase transition, the violation of the chaos bound occurs solely within a stable branch of a small black hole. Moreover, regardless of whether the phase transition takes place, the violations are observed.} 

\section{Introduction}

Phase transitions of black holes (BHs) have always attracted people's attention. The foundational work in this direction was initiated by Hawking and Page through their investigation of the Schwarzschild anti-de Sitter (AdS) spacetime \cite{LP1}. They discovered the first-order phase transition between a large BH and a thermal AdS vacuum, known as the Hawking-Page phase transition. When the temperature is below the critical value, the thermal AdS vacuum phase is more stable; conversely, when the temperature exceeds the critical value, the BH phase dominates. In the thermodynamic context of an extended phase space, where the cosmological constant is regarded as thermodynamic pressure \cite{LP2,LP3}, the phase structure of AdS BHs is fascinating. In \cite{LP4}, Kubiznak and Mann examined the thermodynamic properties of the Reissner-Nordström (RN) AdS BH within the framework of an extended phase space. It was found that this BH exhibits van der Waals-like phase transition behavior, with the P-V criticality matching the P-T diagram, and the P-T coexistence line terminating at the phase transition point. Subsequently, people have conducted in-depth studies on the phase behavior and P-V criticality of other BHs within the extended phase space framework \cite{LP5,LP6,LP7}, and determined the critical exponents associated with phase transitions. These critical exponents are found to be in perfect agreement with those of classical van der Waals fluids. For other interesting results on thermodynamic research in extended phase spaces, please refer to \cite{LP8,LP9,LP10,LP11,LP12,LP13,LP14,LP15,LP16,LP17,LP18,LP19,LP20,LP21,LP22,LP23}.

In recent years, people have begun to explore the connections between black hole (BH) phase transitions and observable physical quantities. For instance, the behaviors of physical quantities such as the circular orbit radius of test particles \cite{LP24,LP25,LP26}, the BH shadow \cite{LP27,LP28}, and quasinormal modes (QNMs) \cite{LP29,LP30,LP31,LP32} can reveal the phase structure of BHs. The discontinuous changes in these physical quantities near the phase transition point are analogous to the behavior of order parameters.

A Lyapunov exponent (LE) serves as an indicator to describe the rate of separation of adjacent trajectories in a dynamical system. A positive exponent value indicates that adjacent orbits diverge over time, signifying a chaotic system, while a negative exponent value implies that adjacent orbits converge over time, describing the stability of the system's motion. It has been widely applied in the study of chaotic dynamics in general relativity \cite{LP33,LP34,LP35,LP36,LP37,LP38,LP39,LP40,LP41,LP42,LP43}. It is not only used to identify the chaotic motion of particles in BH backgrounds but is also closely related to fundamental issues such as BH area quantization and the determination of QNM frequencies. Notably, there is a close relationship between the exponent and the imaginary part of QNMs \cite{LP44}, and BHs' QNMs are associated with their phase transitions. Recent research has provided a new approach for using LEs to detect phase transitions \cite{LP45}. In this work, the authors first investigated the relationship between the chaotic motion of particles around the RN AdS BH and its thermodynamic phase transition. The study revealed that when the phase transition occurs, the exponent exhibits multi-valuedness, with its branches corresponding to the phase structure of this BH. Moreover, the discontinuous change in the exponent can be regarded as an order parameter, possessing a critical exponent of 1/2 near the critical point. Subsequent research has extended this work to other spherically symmetric spacetimes \cite{LP46,LP47,LP48,LP49,LP50,LP51,LPYZD}.

There are various methods for determining the exponent discussed in previous works, \cite{HT,LP73,LP44,LP72,LP74}. In this paper, we adopt the method proposed by Lei and Ge \cite{LP70,LP71}, which determines the exponent by calculating the eigenvalues of the Jacobian matrix. We investigate the LEs of both massless and charged particles orbiting a nonlinear electrodynamics (NLED) BH in the canonical ensemble and explore their relationships with the phase transitions of this BH and the chaos bound. The NLED model, as an extension of the Born-Infeld and Euler-Heisenberg electrodynamics, has garnered significant attention since it was proposed. For example, this model can explain the early universe expansion \cite{LP53}; certain NLED models can serve as alternatives to dark energy to describe the accelerated expansion of the universe and eliminate the Big Bang singularity \cite{LP54}. The first regular BH solution of this model was gotten by Bardeen \cite{LP58}. Subsequently, Bronnikov discovered a class of magnetically charged regular BHs within the framework of coupling general relativity with a specific model \cite{LP59}. Hayward found a specific model that can describe BH collapse and evaporation \cite{LP60}. Balart and Vagenas derived multiple regular BH solutions with nonlinear electrodynamics \cite{LP61}, and Yu and Gao derived an exact solution for the RN BH with nonlinear electromagnetism \cite{LP62}. For regular solutions derived from other NLED models, please refer to \cite{LP68}. 

The rest of this paper is organized as follows. In the next section, we first briefly review the thermodynamics of the NLED BH. In Section III, we calculate the LEs for both massless and charged particles orbiting the NLED BH and explore their relationship with the phase transition. In the fourth section, we discuss the chaos bound in this BH. The last section presents our conclusions.

\section{Thermodynamics of NLED BH}

The minimal interaction between NLED theory and gravity is described as

\begin{eqnarray}
S = \frac{1}{16\pi} \int \sqrt{-g} \left( R + K(\psi) \right)  d^4x,
\label{eq2.1.1}
\end{eqnarray}

\noindent where

\begin{eqnarray}
F_{\mu\nu} = \nabla_\mu A_\nu - \nabla_\nu A_\mu, \quad \psi = F_{\mu\nu} F^{\mu\nu},
\label{eq2.1.2}
\end{eqnarray}

\noindent In the above equation, $R$ represents the Ricci scalar, $A_\nu $ denotes the Maxwell field, and $K(\psi)$ is a nonlinear function of $\psi$. The corresponding field equations are obtained by varying the above action.

\begin{eqnarray}
G_{\mu\nu} = \frac{1}{2} g_{\mu\nu} K[\psi] - 2K[\psi]_{,\psi} F_{\mu\lambda} F^{\lambda}_{\nu},
\label{eq2.1.3}
\end{eqnarray}

\noindent where $K[\psi]_{,\psi} = \frac{dK[\psi]}{d\psi}$, the corresponding generalized Maxwell equations are expressed in the following form

\begin{eqnarray}
\nabla_{\mu}(K_{,\psi} F^{\mu\nu}) = 0.
\label{eq2.1.4}
\end{eqnarray}

The spherically symmetric static metric is given by \cite{LP62}

\begin{eqnarray}
ds^2 = -F(r)dt^2 + \frac{1}{F(r)}dr^2 +r^2d\theta^2+r^2{\sin^2\theta}d\phi^2,
\label{eq2.1}
\end{eqnarray}

\noindent The only non-zero component of the Maxwell field tensor is $A_t = -\phi(r)$, and $\psi = -2\varphi'^2$. Next, we consider the specific expression $K[\psi] = -\psi - 2\Lambda + 2\sqrt{2}\alpha(-\psi)^{\frac{1}{2}}$, where $\alpha$ is coupling constant and $\Lambda$ is cosmological constant factor. The corresponding solution to the field equayions is given by \cite{LP62}   

\begin{eqnarray}
\phi(r) = \frac{Q}{r} + r\alpha,
\label{eq2.1.5}
\end{eqnarray}
\begin{eqnarray}
F(r) = 1 - \frac{2M}{r} + \frac{Q^2}{r^2} + \frac{r^2}{l^2} + 2{\alpha}Q - \frac{{\alpha}^2 r^2}{3},
\label{eq2.3}
\end{eqnarray}

\noindent \noindent $Q$  and $\alpha$ are the charge, coupling constant, respectively. And $l$ is the  AdS radius, in the extended phase space, the cosmological constant is treated as pressure, defined as $P = \frac{-\Lambda}{8\pi} = \frac{3}{8\pi l^2}$.

\begin{figure}[H]
  \centering
  \begin{subfigure}[b]{0.48\textwidth}
    \includegraphics[width=6.5cm,height=6cm]{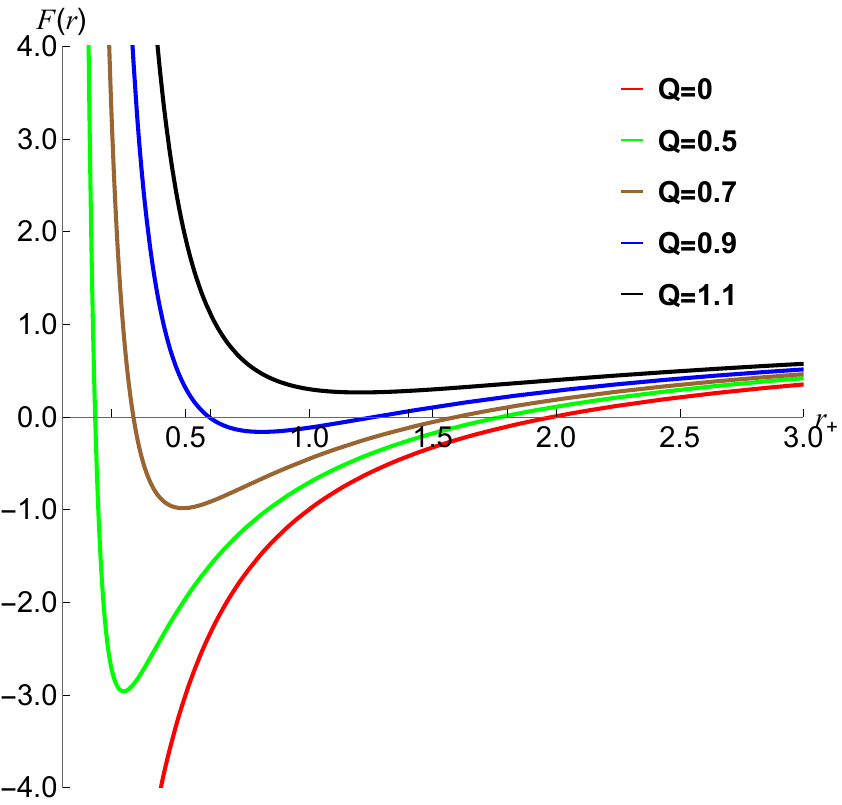}
    \caption{$\alpha=0.04$ and $M=1$}
    \label{f21-a}
  \end{subfigure}
  \hfill 
  \begin{subfigure}[b]{0.48\textwidth}
    \includegraphics[width=6.5cm,height=6cm]{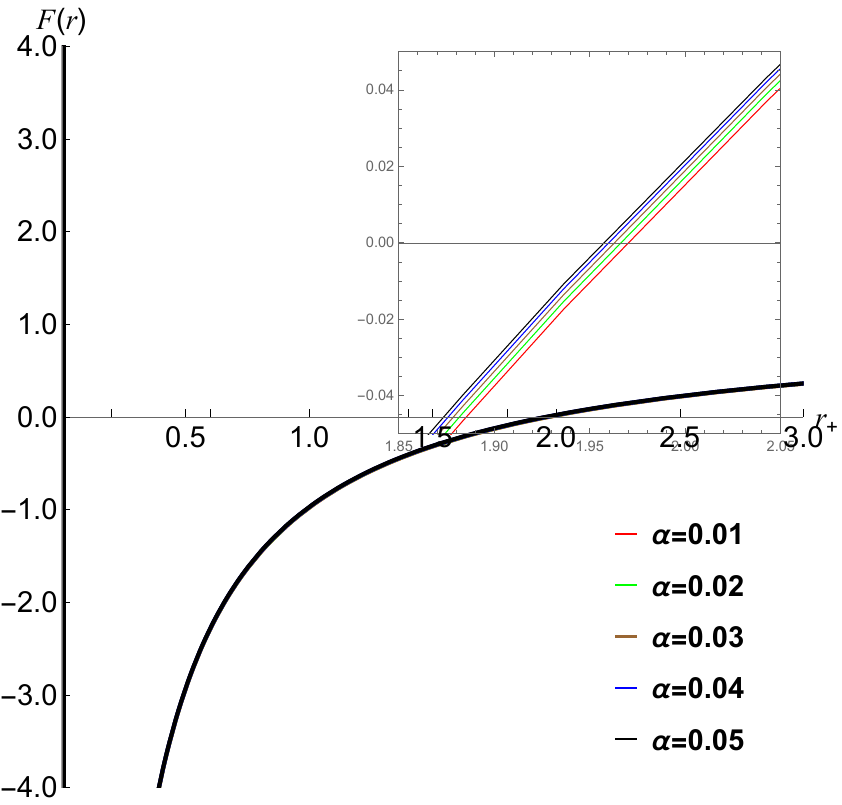}
    \caption{$Q=0.12$ and $M=1$}
    \label{f21-b}
  \end{subfigure}
  \caption{The variation of the F(r) with respect to the r of the NLED BH }
  \label{f21}
\end{figure}

Using Eq. (\ref{eq2.3}), the physical properties of the black hole event horizon can be determined when $F(r)=0$, Figure (\ref{f21}) shows the behavior of the event horizon for different values of $Q$ and $\alpha$. From Figure \ref{f21-a}, it is seen that the size of the black hole increases with the increase of the charge $Q$. When $Q=1.1$, a naked singularity appears, which means that no black hole exists when $Q$ exceeds the threshold. When $Q$ is smaller than the threshold, two simple zeros appear, corresponding to two event horizons, as shown in Figure \ref{f21-a} for $Q=0.5,0.7,0.9$. In Figure \ref{f21-b}, it is clear that the size of the black hole is also affected by the coupling constant $\alpha$, and and the size of the black hole increases as $\alpha$ increases.

 The event horizon is located at $r=r_+$ and determined by $F(r)=0$.  $M$ is the mass and can be expressed in terms of the horizon as  

\begin{eqnarray}
M = \frac{r_+}{2} \left( 1 + \frac{Q^2}{r_+^2} + \frac{r_+^2}{l^2} + 2{\alpha}Q - \frac{{\alpha}^2 r_+^2}{3} \right). 	
\label{eq2.4}
\end{eqnarray}

\noindent When $\alpha=0$, the metric is reduced to the RN AdS metric. When $\alpha=Q=0$, the metric recovers the Schwarzschild metric. The Hawking temperature and entropy are \cite{LP69}

\begin{eqnarray}
T &=& \frac{\kappa}{2\pi}= \frac{1}{4\pi} \biggl( \frac{1}{r_+} + \frac{3 r_+}{l^2} - \frac{Q^2}{r_+^3} + \frac{2\alpha Q}{r_+} - {\alpha}^2 r_+ \biggr),\label{eq2.5} \\
S &=& \pi{r_+^2}.
\label{eq2.5.1}
\end{eqnarray}

The Gibbs free energy is

\begin{eqnarray}
G = M - T S = \frac{1}{12r_+} (9Q^2 + 3{r_+}^2 + 6{\alpha}Q{r_+}^2 + ({\alpha}^2 - \frac{3}{l^2}) {r_+}^4).
\label{eq2.6}
\end{eqnarray}

\noindent By using Eq. (\ref{eq2.5}), we can demonstrate that the horizon is a function of the temperature. When a specific temperature value is correlated with multiple values of the horizon radius, it serves as a clear indication that the BH experiences multiple distinct phases at that particular temperature. On the contrary, in the scenario where such a multi-valued correlation does not exist, it implies that no phase transition takes place within the BH. It is noteworthy that the critical temperature for this phase transition is determined by

\begin{eqnarray}
\frac{\partial {T}}{\partial {r}_+} = 0, \quad \frac{\partial^2 {T}}{\partial {r}_+^2} = 0.
\label{eq2.7}
\end{eqnarray}
 
\begin{figure}[H]
	\centering
	\includegraphics[width=8cm,height=7cm]{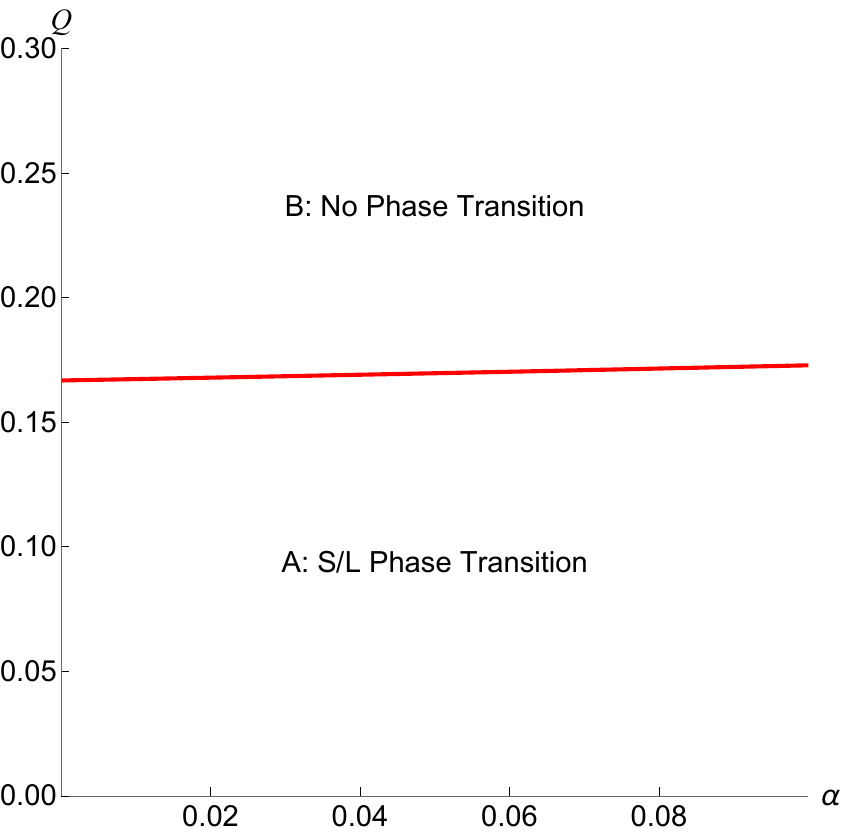}
	\caption{The critical curve in $Q-\alpha$ parameter space.}
	\label{f22}
\end{figure}

 \noindent The critical curve in the coupling constant and charge ($Q-\alpha$) space is shown in Figure \ref{f22}. The curve divides the plane into two distinct regions. Region A corresponds to the parameter range in which the BH undergoes a S/L phase transition, while in Region B the BH remains in a stable single phase with no phase transition.

\noindent  From the above equation, we obtained the analytical solution at the critical point,

\begin{eqnarray}
{r}_{+c} = \sqrt{\frac{9 l^2 - 3 a^2 l^4 + \sqrt{9 a^2 l^6 - 3 a^4 l^8}}{54 - 42 a^2 l^2 + 8 a^4 l^4}}, \quad  {Q}_c=\frac{a^2 l^4 + \sqrt{9 a^2 l^6 - 3 a^4 l^8}}{2 a l^2 \left( 9 - 4 a^2 l^2 \right)},
\label{eq2.1.6}
\end{eqnarray}

\noindent and

\begin{eqnarray}
T_c = \frac{(3 - a^2l^2) \sqrt{\frac{18l^2 - 6a^2l^4 + 2\sqrt{9a^2l^6 - 3a^4l^8}}{27 - 21a^2l^2 + 4a^4l^4}}}{3l^2\pi}.
\label{eq2.1.7}
\end{eqnarray}

\noindent We order ${\alpha}=0.04$, $l=1.00$,  and get the values at the critical point as follows 

\begin{eqnarray}
{r}_{+c}=0.411108,\quad  {Q}_c=0.168965,  \quad {T}_c=0.272423.
\label{eq2.8}
\end{eqnarray}

\begin{figure}[h]
	\centering
	\includegraphics[width=10cm,height=7cm]{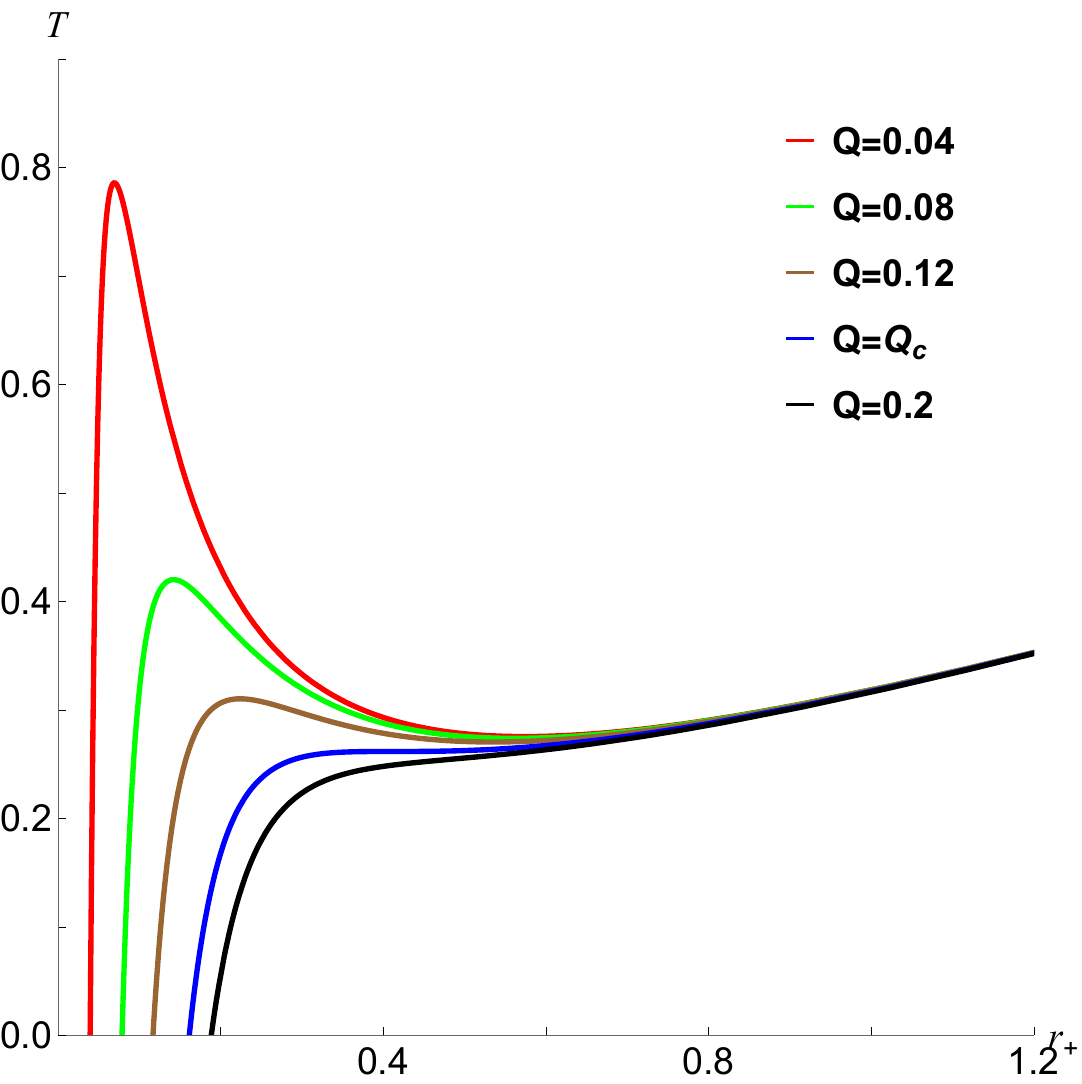}
	\caption{The variation of the temperature with respect to the horizon radius of the NLED BH, where ${\alpha}=0.04$.}
	\label{f1}
\end{figure}

\noindent A graphical representation of the temperature as a function of the  horizon radius is presented in Figure \ref{f1}. Each curve depicted in this figure corresponds to a distinct value of the BH's charge. The figure reveals that no maximum values are present, which provides an evidence that the BH exists in a single phase state under the condition of ${Q}>{Q_c}$. Conversely, when ${Q}<{Q_c}$, maximum values become evident in the plot. This phenomenon implies that the BH undergoes a transition to a multi phase state, with multiple distinct phases coexisting within its structure.

\begin{figure}[h]
  \centering
  \begin{subfigure}[b]{0.48\textwidth}
    \includegraphics[width=6.5cm,height=6cm]{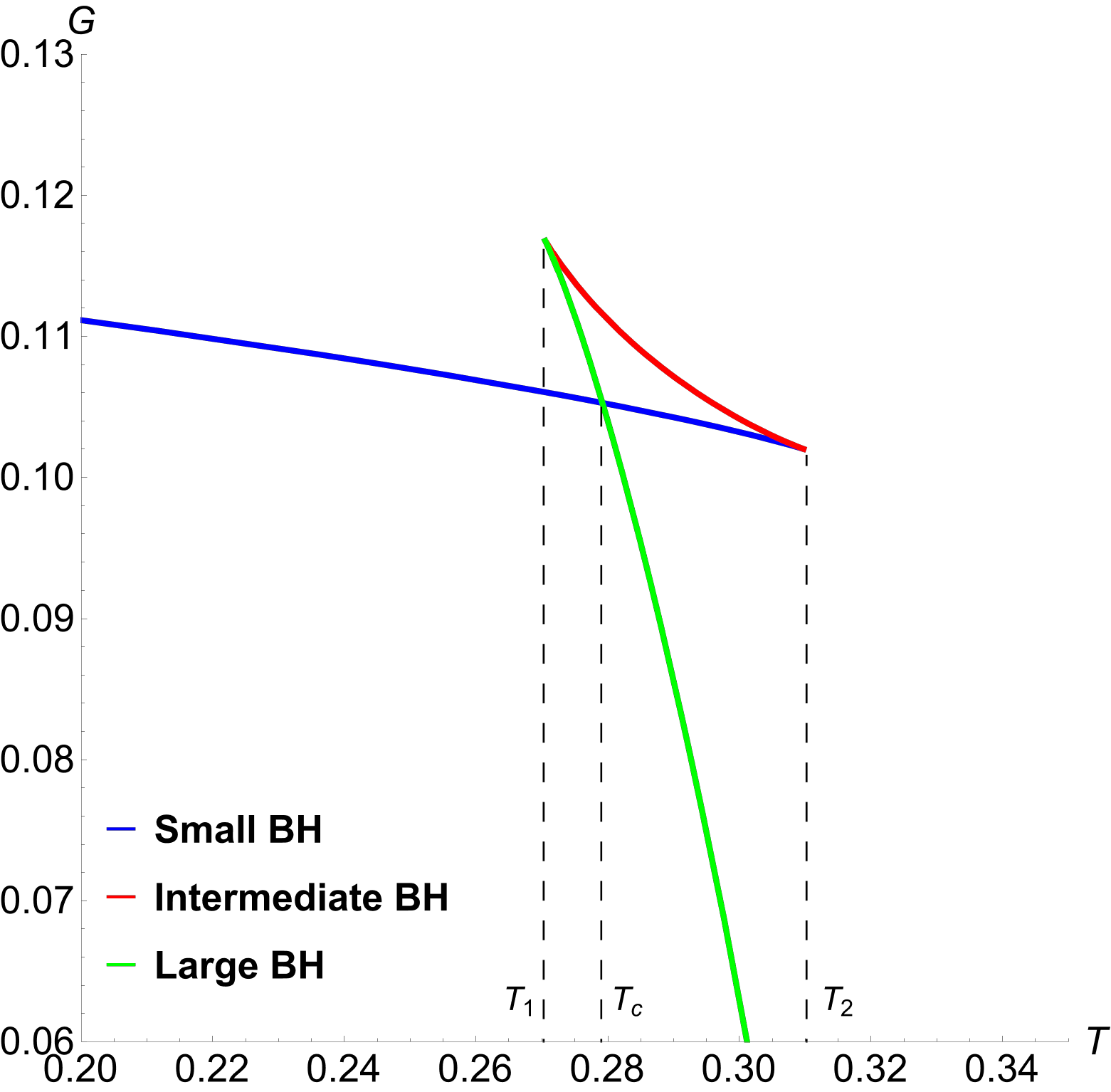}
    \caption{${Q}=0.12<{Q}_c$}
    \label{f2-a}
  \end{subfigure}
  \hfill 
  \begin{subfigure}[b]{0.48\textwidth}
    \includegraphics[width=6.5cm,height=6cm]{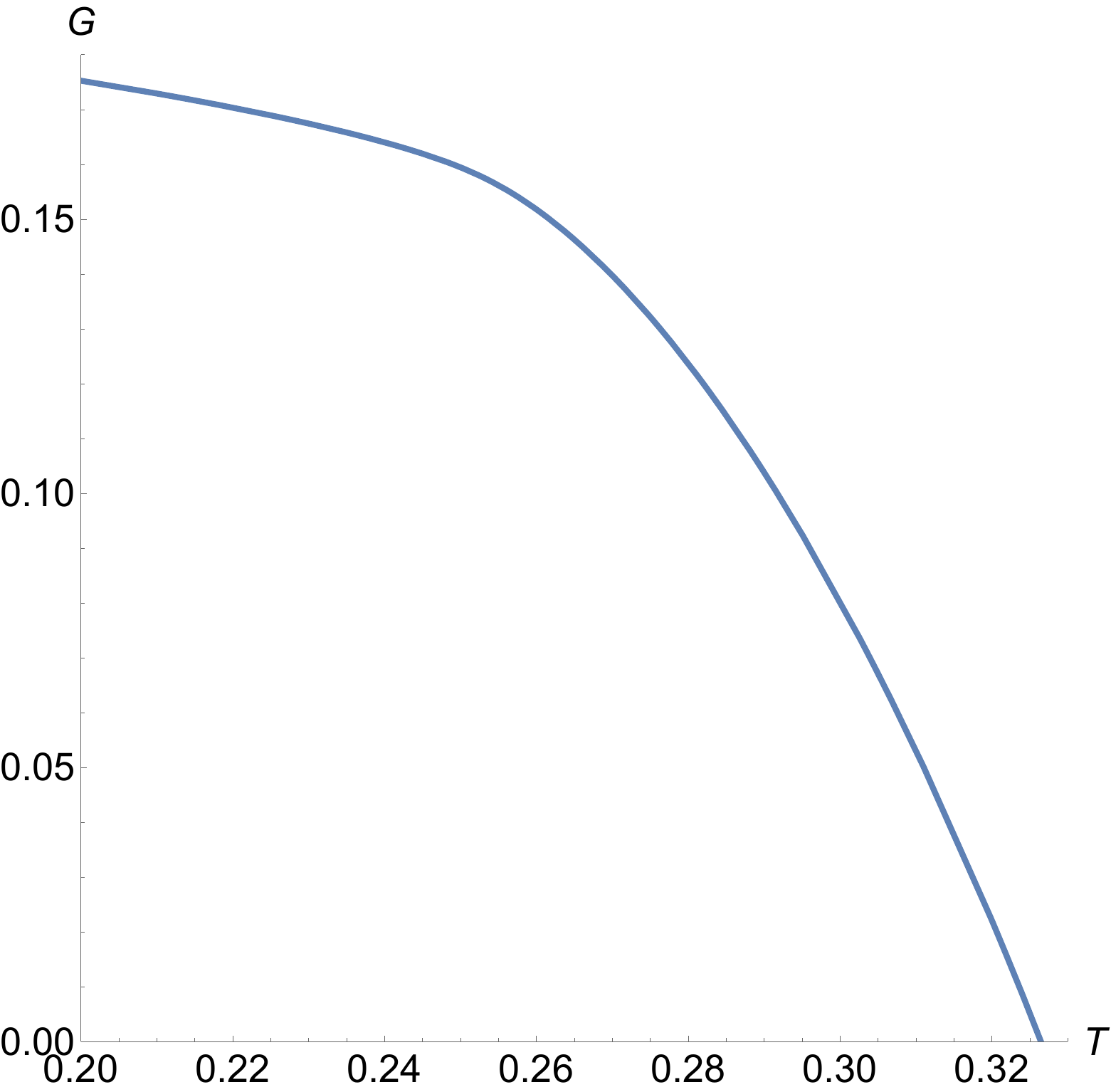}
    \caption{${Q}=0.20>{Q}_c$}
    \label{f2-b}
  \end{subfigure}
  \caption{The variation of the Gibbs free energy with respect to the temperature of the NLED BH, where ${\alpha}=0.04$. }
  \label{f2}
\end{figure}

To further study the phase transition, we utilized Eq. (\ref{eq2.5}) and Eq. (\ref{eq2.6}) to establish the relationship between the Gibbs free energy and the temperature. The corresponding graphical representation is presented in Figure \ref{f2}. As clearly illustrated in Figure \ref{f2}, when ${Q}<{Q_c}$, multiple phases emerge. These phases can be precisely categorized into three types: small BH, intermediate BH, and large BH. It is of great significance to note that these three BH phases coexist within the temperature range ${T_1}<{T}<{T_2}$. The temperature ${T_c}$ marks the point where a first-order phase transition occurs. Conversely, when ${Q}>{Q_c}$, the BH system does not undergo any phase transition.

\section{LEs and phase transition of NLED BH}

In this section, we first obtain the LEs of chaos for the massless and charged particles in the equatorial plane around the four-dimensional NLED BH, and then explore their relationships with the phase transition. There are lots of work to derived the LEs. First, we review the method developed in \cite{LP70,LP71}.

\subsection{LEs }\label{sec3.1}

When a particle with charge $q$ moves in a circular orbit within the equatorial plane of the BH, its Lagrangian is

\begin{eqnarray}
\mathcal{L} = \frac{1}{2}\left(-F\dot{t}^2+\frac{\dot{r}^2}{F} +r^2\dot{\phi}^2\right) -qA_t\dot{t},
\label{eq2.9}
\end{eqnarray}

\noindent where $\dot{x} = \frac{d}{d\tau}$ and $\tau$ is a proper time. From the Lagrangian and using the definition of generalized momenta $ \pi_{\mu}= \frac{\partial\mathcal{L}}{\partial\dot{x}}$, we get

\begin{eqnarray}
\pi_t = \frac{\partial\mathcal{L}}{\partial\dot{t}} = -F\dot{t} -qA_t=-E, \quad \pi_r = \frac{\partial\mathcal{L}}{\partial\dot{r}} = \frac{\dot{r}}{F}, \quad \pi_{\phi} =\frac{\partial\mathcal{L}}{\partial\dot{\phi}} = r^2  \dot{\phi}=L.
\label{eq2.10}
\end{eqnarray}

\noindent In the above equation,  $E$ and $L$ are the energy and angular momentum of the particle, respectively. Thus the Hamiltonian is

\begin{eqnarray}
H = \frac{1}{2F}({-(\pi_{t}+qA_{t})^2+\pi_r^2F^2+ \pi^2_{\phi}r^{-2}F}).
\label{eq2.11}
\end{eqnarray}

\noindent From the Hamiltonian, the equations of motion are

\begin{eqnarray}
\dot{t} &=& \frac{\partial H}{\partial \pi_t}=-\frac{\pi_t+qA_t}{F}, \quad  \dot{\pi_t}= -\frac{\partial H}{\partial t} =0 ,
\quad \dot{r} = \frac{\partial H}{\partial \pi_r}= \pi_r F, \nonumber\\
\dot{\pi_r} &=& -\frac{\partial H}{\partial r} =-\frac{1}{2}\left[\pi^2_r F^{\prime} -\frac{2qA^{\prime}_t(\pi_t+qA_t)}{F}+\frac{(\pi_t+qA_t)^2F^{\prime}}{F^2} -\pi^2_{\phi}(r^{-2})^{\prime}\right], \nonumber\\
\dot{\phi} &=& \frac{\partial H}{\partial \pi_{\phi}}= \frac{\pi_{\phi}}{r^2}, \quad  \dot{\pi_\phi}= -\frac{\partial H}{\partial \phi} =0.
\label{eq2.12}
\end{eqnarray}

\noindent In the above equations, the symbol "$\prime$" represents the derivative with respect to $r$. The normalization of the four-velocity of a particle is given by $ g_{\mu\nu}\dot{x}^{\mu}\dot{x}^{\nu}=\eta$, where $\eta =0$ describes the case of a massless particle, and $\eta =-1$ corresponds to the case of a massive particle. Using the normalization and the metric (\ref{eq2.1}), we get a constrain condition for the charged particle,

\begin{eqnarray}
\pi_t+qA_t=-\sqrt{F(1+ \pi_r^2F + \pi_{\phi}^2r^{-2})},
\label{eq2.13}
\end{eqnarray}

\noindent Using this constrain, we obtain radial motion equations at time $t$,

\begin{eqnarray}
\frac{dr}{dt} &=& \frac{\dot{r}}{\dot{t}} =-\frac{\pi_rF^2}{\pi_t+qA_t}, \nonumber\\
\frac{d\pi_r}{dt} &=& \frac{\dot{\pi_r}}{\dot{t}} =  -qA'_t + \frac{1}{2}\left[\frac{\pi_r^2FF'}{\pi_t+qA_t}+\frac{(\pi_t+qA_t)F'}{F}-\frac{\pi_\phi^2(r^{-2})'F}{\pi_t+qA_t}\right].
\label{eq2.14}
\end{eqnarray}

\noindent In order to calculate the LE, we select a phase space $(r, \pi_r)$ and define a matrix with its matrix elements given by

\begin{eqnarray}
K_{11} = \frac{\partial F_1}{\partial r}, \quad
K_{12} = \frac{\partial F_1}{\partial \pi_r}, \quad
K_{21} = \frac{\partial F_2}{\partial r}, \quad
K_{22} = \frac{\partial F_2}{\partial \pi_r}.
\label{eq2.15}
\end{eqnarray} 

\noindent For convenience, we have defined $F_1= \frac{dr}{dt}$ and $F_2=\frac{d\pi_r}{dt}$. Taking into account the motion of the particle in an equilibrium orbit, we get $\pi_r=\frac{d\pi_r}{dt}=0$. From the constraint Eq. (\ref{eq2.13}), the  matrix elements can be rewritten as \cite{LP75}

\begin{eqnarray}
K_{11} &=& \frac{\partial F_1}{\partial r}=0, \quad K_{22} =\frac{\partial F_2}{\partial \pi_r}=0, \nonumber\\
K_{12}&=& \frac{\partial F_1}{\partial \pi_r} =\frac{F^2}{\sqrt{F(1+ \pi_{\phi}^2r^{-2})}},\nonumber\\
K_{21}&=& \frac{\partial F_2}{\partial _r} = -qA^{\prime\prime}_t
-\frac{F^{\prime\prime}+ \pi^2_{\phi}(r^{-2}F)^{\prime\prime}}{2\sqrt{F(1+ \pi_{\phi}^2r^{-2})}} + \frac{\left[F^{\prime} +\pi_{\phi}^2(r^{-2}F)^{\prime}\right]^2}{4\left[F(1 + \pi_{\phi}^2r^{-2})\right]^{3/2}}.
\label{eq2.16}
\end{eqnarray}

 \noindent By calculating the eigenvalues of the above matrix, we obtain the Lyapunov exponent at the specific radial position of the circular orbit, which takes the following form 
\begin{eqnarray}
\lambda^2 = \frac{1}{4} \left[ \frac{F' + \pi_{\phi}^2 (r^{-2} F)'}{1 + \pi_{\phi}^2 r^{-2}} \right]^2 - \frac{1}{2} F\frac{F'' + \pi_{\phi}^2 (r^{-2} F)''}{1 + \pi_{\phi}^2 r^{-2}} - \frac{q A_t'' F^2}{\sqrt{F(1 + \pi_{\phi}^2 r^{-2})}}.
\label{eq2.17}
\end{eqnarray}

\noindent It is clearly that the exponent is affected by the angular momentum $\pi_{\phi}=L$, the particle's charge and the BH's charge. 

For a massless particle, its charge is zero. The exponent of chaos for this particle is derived by following a similar procedure analogous to that described above, which is 

\begin{eqnarray}
\lambda^2 = -\frac{1}{2}\frac{F(r^{-2}F)^{\prime\prime}}{r^{-2}}+ \frac{1}{4}\frac{F\left[(r^{-2}F)^{\prime}\right]^2}{F(r^{-2})^2}.
\label{eq2.18}
\end{eqnarray}

\noindent Obviously, it is not affected by the angular momentum at this time. For the derivation of the exponent for massless particles, please refer to the Appendix.

\subsection{Massless particle's case}\label{sec3.2}

\begin{figure}[h]
  \centering
  \begin{subfigure}[b]{0.48\textwidth}
    \includegraphics[width=7cm,height=5.5cm]{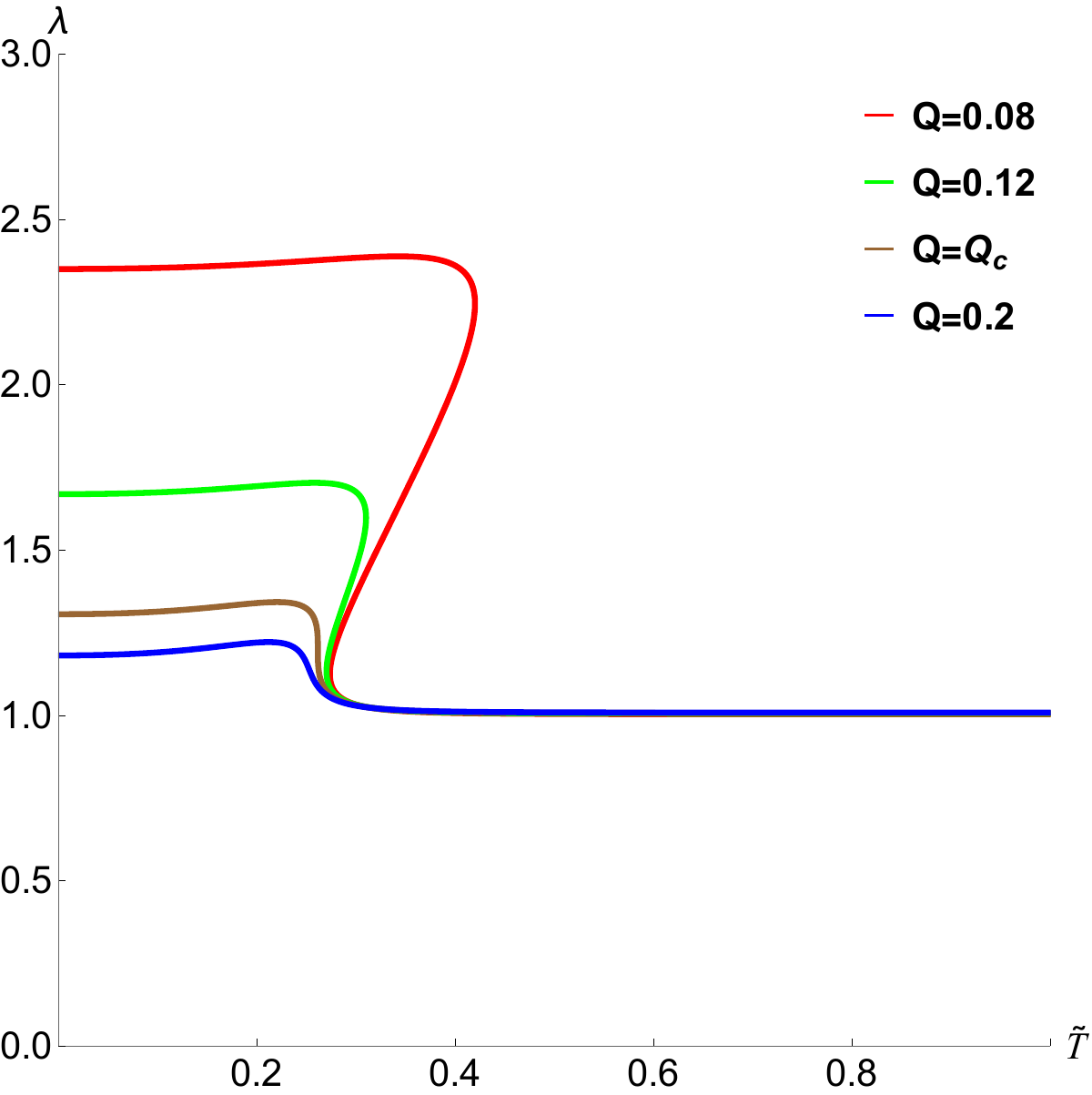}
    \caption{LE vs the temperature}
    \label{f3-a}
  \end{subfigure}
  \hfill 
  \begin{subfigure}[b]{0.48\textwidth}
    \includegraphics[width=7cm,height=5.5cm]{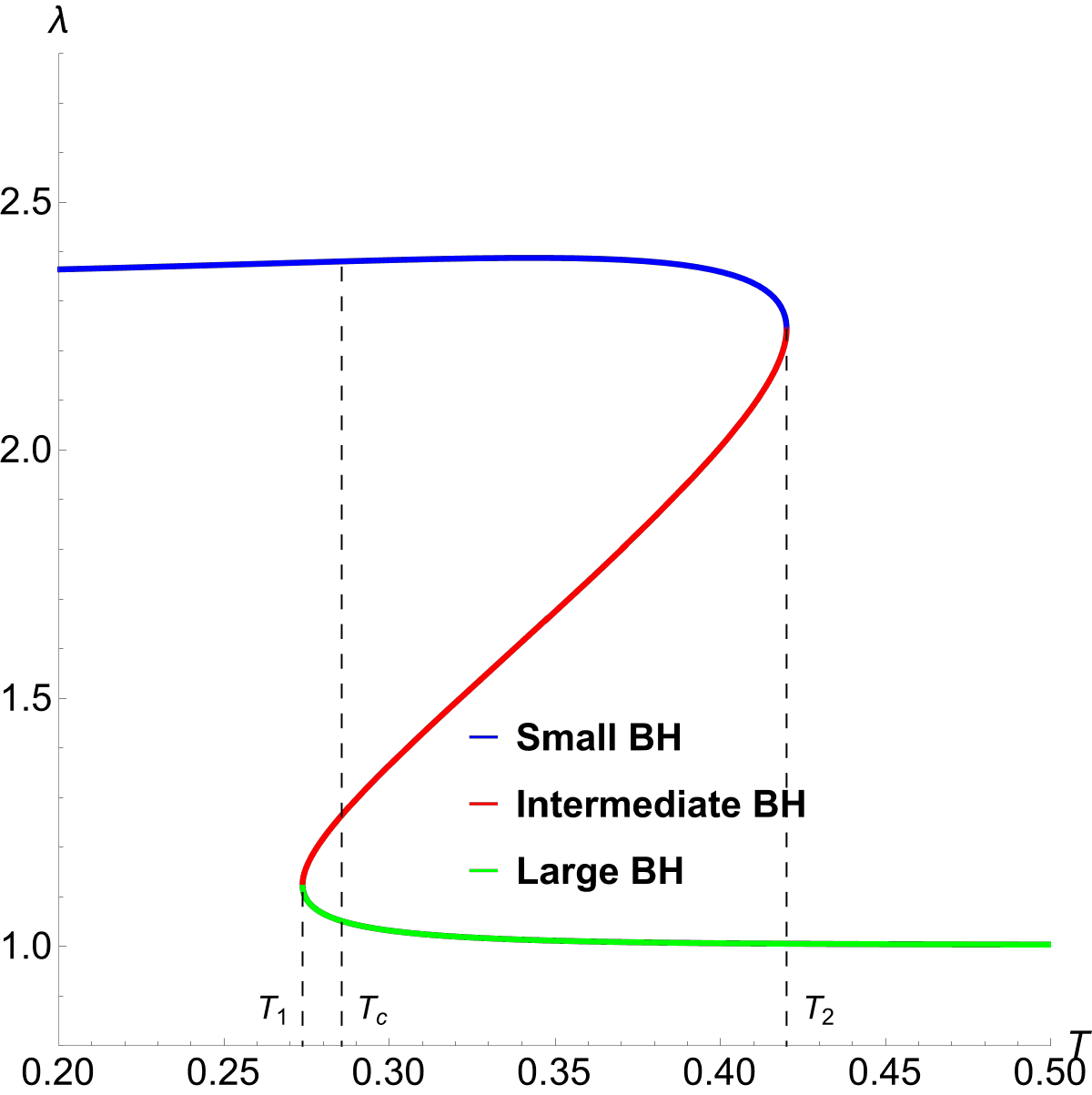}
    \caption{${Q}=0.12$}
    \label{f3-b}
  \end{subfigure}
  \begin{subfigure}[b]{0.48\textwidth}
    \includegraphics[width=7cm,height=5.5cm]{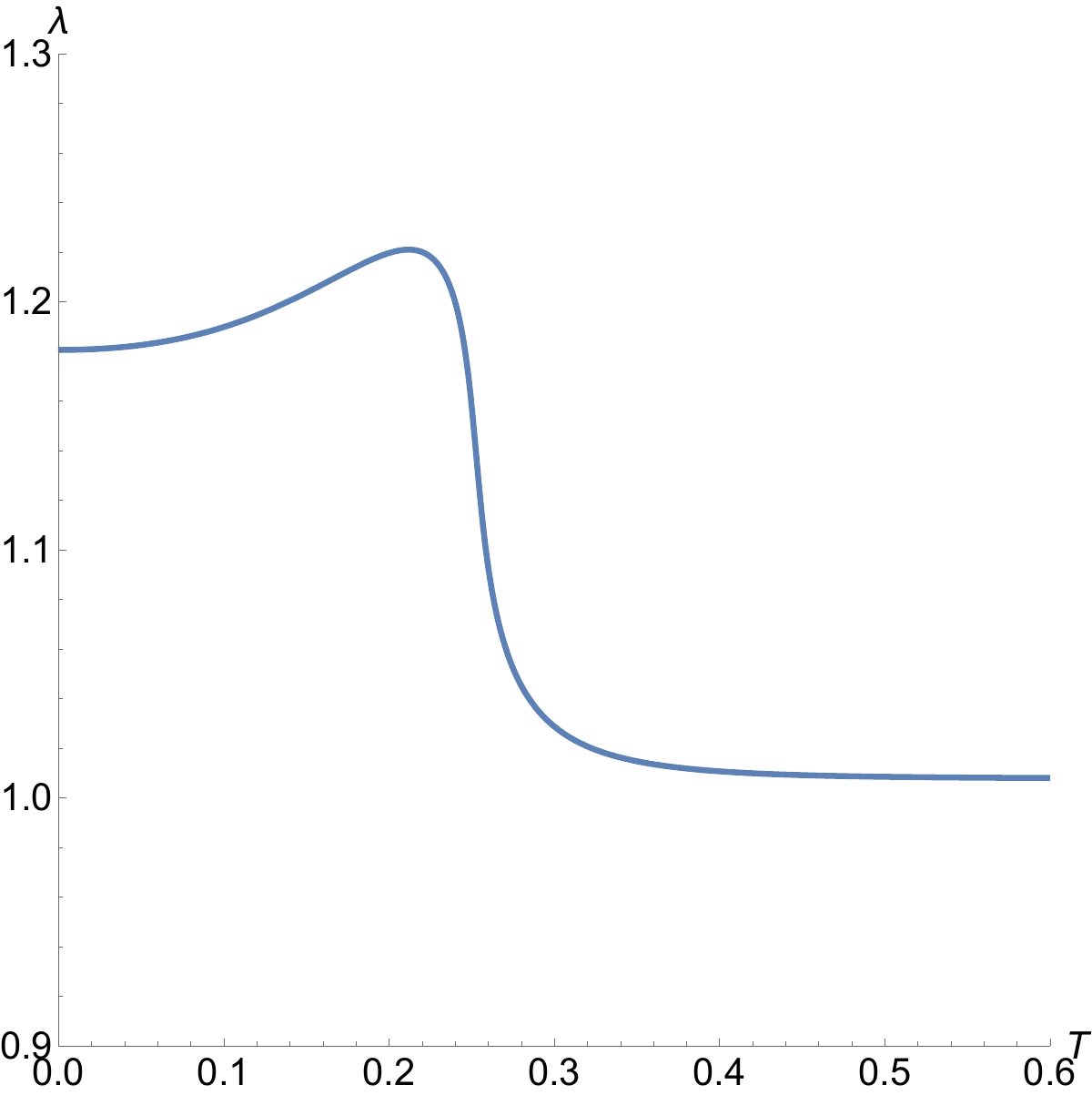}
    \caption{${Q}=0.20$}
    \label{f3-c}
  \end{subfigure}
 \hfill 
  \begin{subfigure}[b]{0.48\textwidth}
    \includegraphics[width=7cm,height=5.5cm]{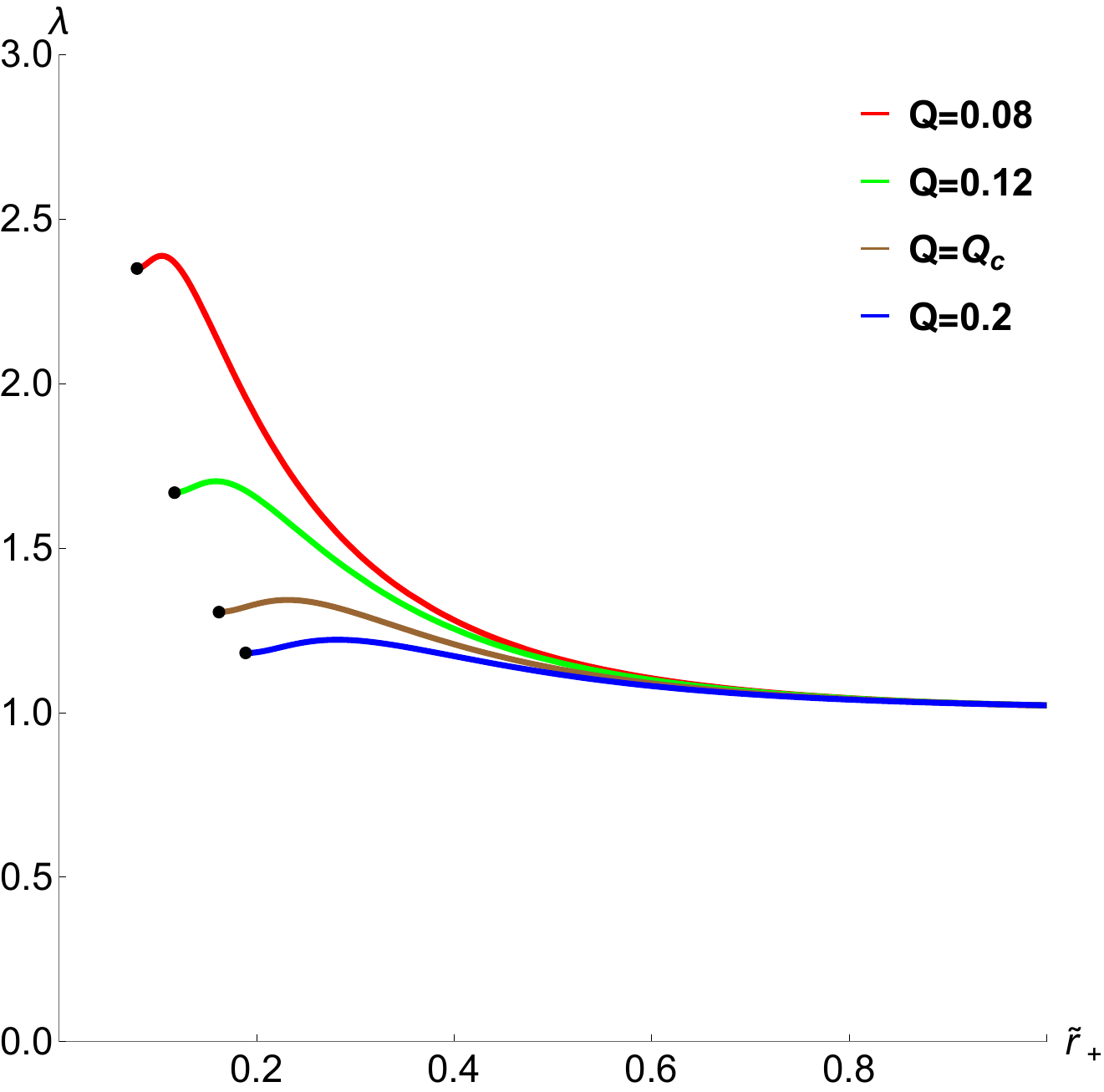}
    \caption{LE vs the horizon radius}
    \label{f3-d}
  \end{subfigure}
  \caption{The variation of the LE of chaos for the massless particle with respect to the temperature of the NLED BH.}
  \label{f3}
\end{figure}

As demonstrated in \cite{LP46,LP47,LP48,LP49,LP50,LP51}, the LEs exhibit remarkable capability in revealing the thermodynamic phase transitions of BHs in AdS spacetimes. Building on this foundation, we first explore the correlation between the exponent of chaos for the massless particle and the phase transition of the NLED BH, and draw Figure \ref{f3}.

In this figure, when ${Q}>{Q}_c$, the exponent is single valued at all temperatures, indicating the absence of the phase transition. When ${Q}<{Q}_c$, the exponent is multivalued, and as the temperature rises, it approaches a constant value. Specifically, Figure \ref{f3-b} illustrates the temperature dependence of the exponent for $Q=0.12$. The exponent is single valued when ${T}<{T}_1$ and ${T}>{T}_2$, while it is multivalued when ${T}_1<{T}<{T}_2$. Within this temperature range ${T}<{T}<{T}_2$, two distinct scenarios emerge: For the small BH branch, the exponent slightly increases and then declines with rising the temperature. For the large BH branch, the exponent decreases with increasing the temperature. In contrast, the intermediate BH branch exhibits an increase in the exponent as the temperature rises. By comparing Figure \ref{f2} and Figure \ref{f3}, we observe that the temperature range with multivalued exponents corresponds precisely to the phase transition region shown in Figure \ref{f2}. Notably, for the large BH branch when  ${T}>{T}_2$, as the temperature keeps increasing, the exponent approaches 1.

For ${Q}>{Q}_c$, the exponent is single valued and approaches 1 as the temperature increases, as demonstrated in Figure \ref{f3-c}. A comparative analysis of Figure \ref{f3-b} and Figure \ref{f3-c} reveals that the multivalued nature of the exponent exclusively emerges when ${Q}<{Q}_c$, showing remarkable consistency with the free energy behavior observed in Figure \ref{f2}. To further elucidate the correlation between the exponent and phase transition, as well as to validate the completeness of our results, Figure \ref{f3-d} presents a graph showing the functional relationship between the exponent and the horizon radius. The black dot points describe the case for ${T}=0$. We observe that the exponent value decreases with increasing the charge at the extremal horizon (the horizon radius located at ${T}=0$). Furthermore, as the horizon radius expands, all exponents approach 1 regardless of the charge's values. This establishes that for the massless particle, the LE serve as an effective probe for detecting the phase transition.

\subsection{Charged particle's case}\label{sec3.3}

\begin{figure}[h]
  \centering
  \begin{subfigure}[b]{0.48\textwidth}
    \includegraphics[width=7cm,height=5.5cm]{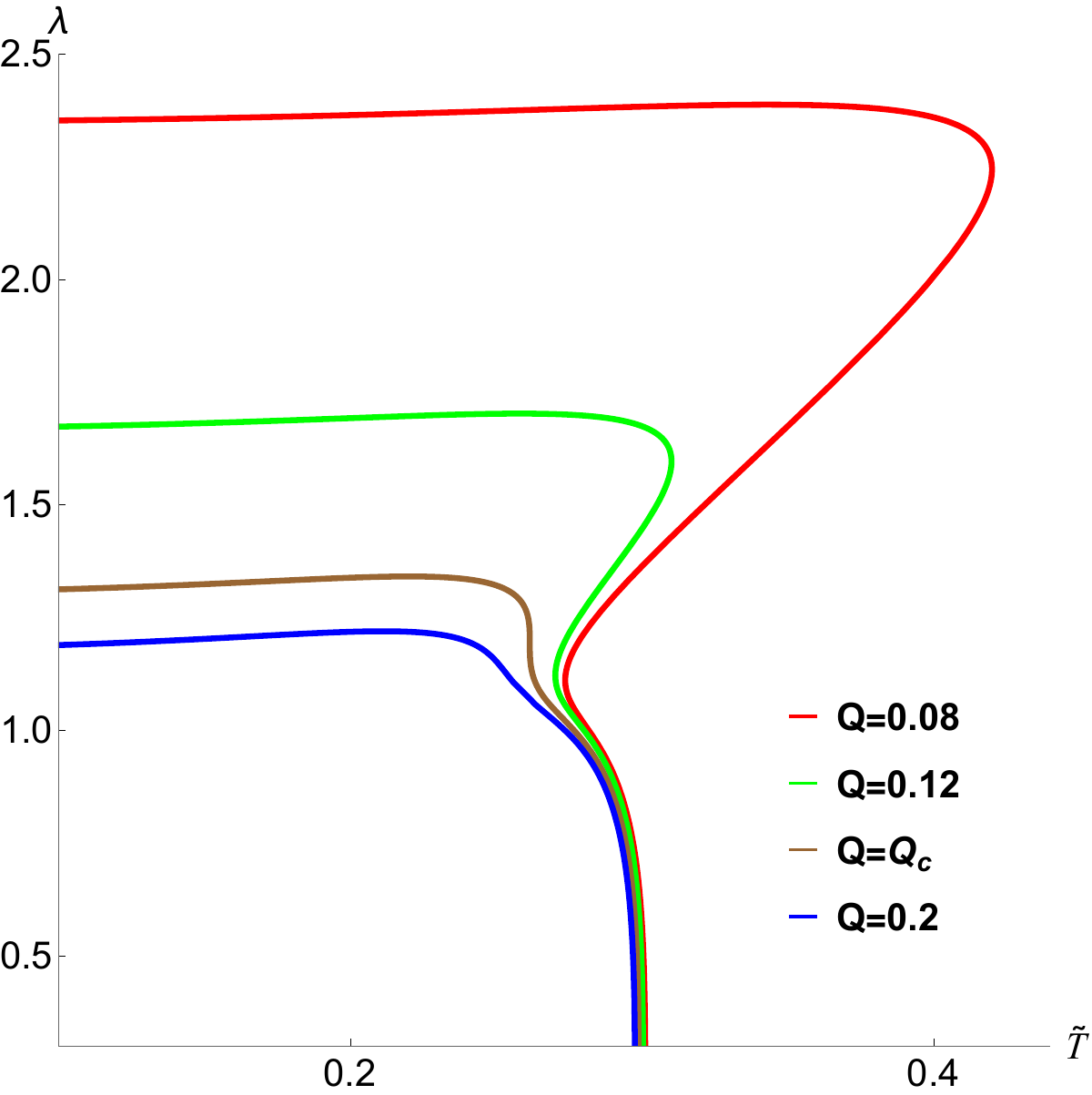}
    \caption{LE vs the temperature}
    \label{f4-a}
  \end{subfigure}
  \hfill 
  \begin{subfigure}[b]{0.48\textwidth}
    \includegraphics[width=7cm,height=5.5cm]{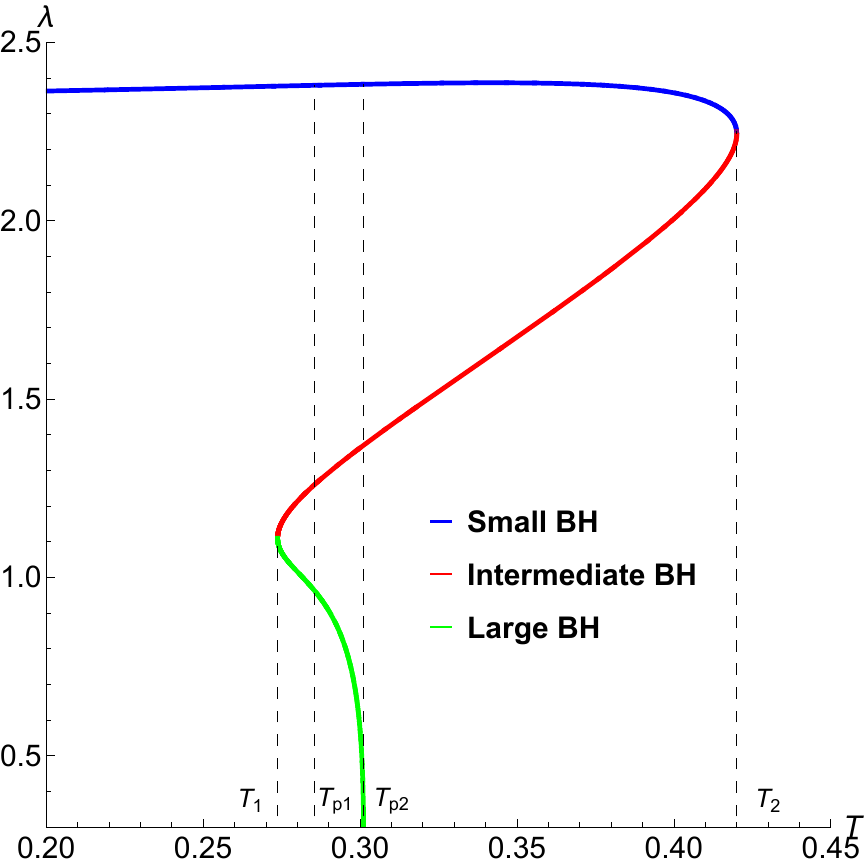}
    \caption{${Q}=0.12$}
    \label{f4-b}
  \end{subfigure}
  \begin{subfigure}[b]{0.48\textwidth}
    \includegraphics[width=7cm,height=5.5cm]{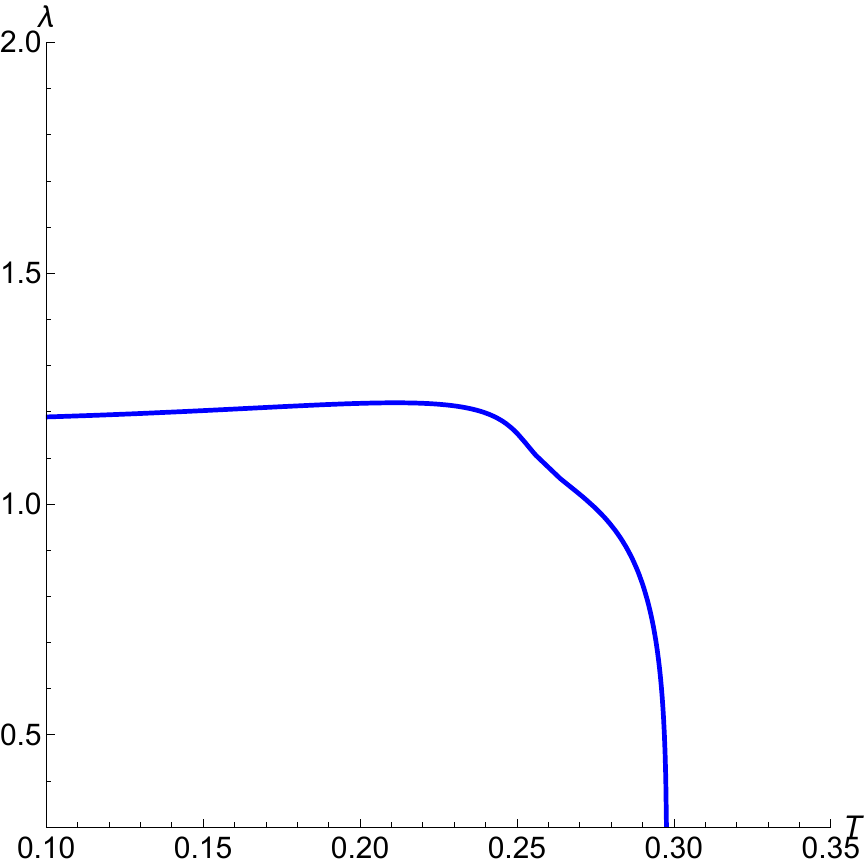}
    \caption{${Q}=0.20$}
    \label{f4-c}
  \end{subfigure}
 \hfill 
  \begin{subfigure}[b]{0.48\textwidth}
    \includegraphics[width=7cm,height=5.5cm]{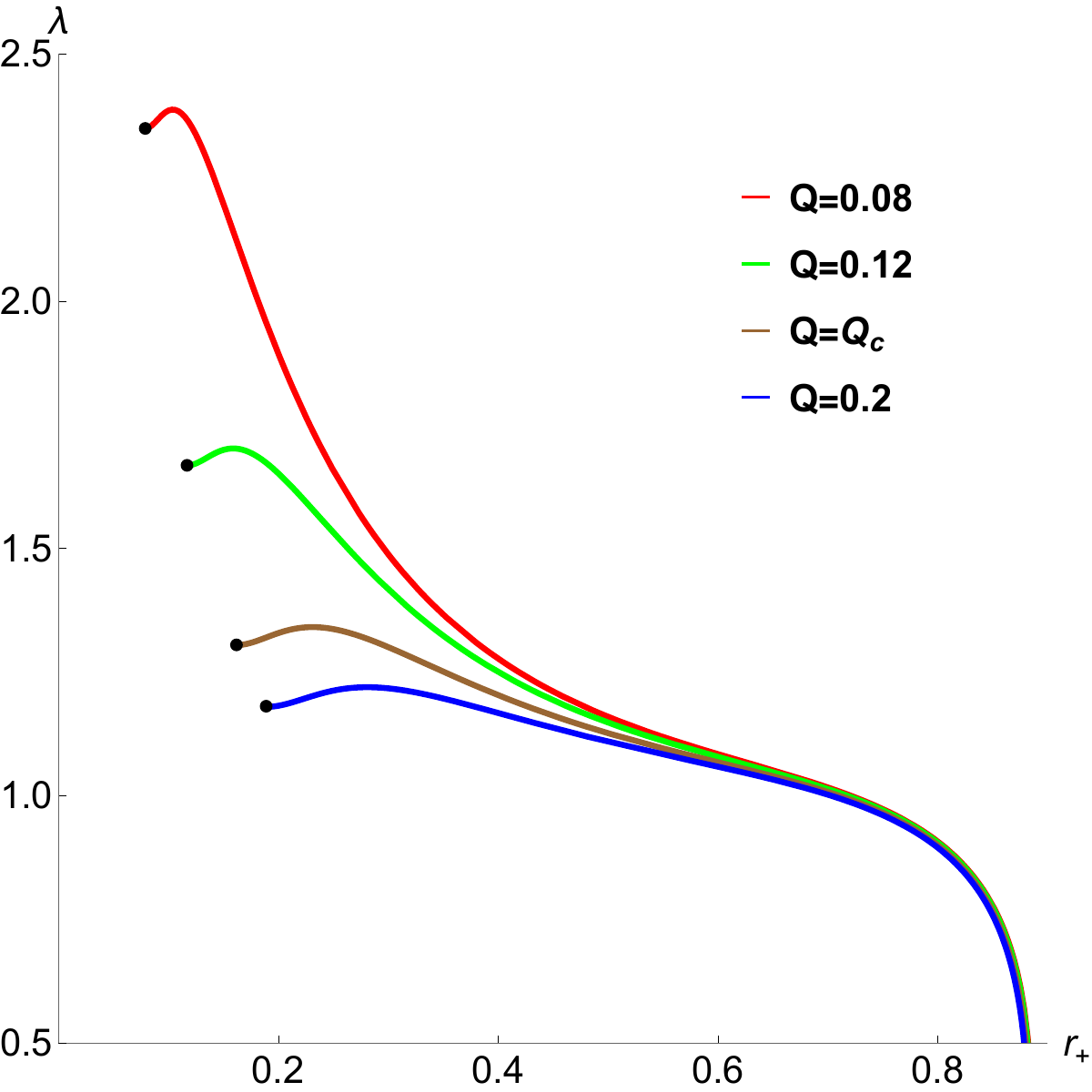}
    \caption{LE vs the horizon radius}
    \label{f4-d}
  \end{subfigure}
  \caption{The variation of the LE of chaos for the charged particle with respect to the temperature of the NLED BH, where $q=0.01$.}
  \label{f4}
\end{figure}

Throughout the calculations in this subsection, we fix $L = 20.00$ and ${\alpha}=0.04$, and for a charged particle, we select $\eta =-1$ during the normalization of its four velocity. The position of its unstable orbit is determined by Eq. (\ref{eq2.14}). By employing Eqs. (\ref{eq2.3}) and (\ref{eq2.17}), we derive the relationship between the exponent and temperature, which is plotted in Figure \ref{f4}. In this analysis, we set $q=0.01$. The results presented in this figure exhibit similarities to those in Figure \ref{f3}. Specifically, when ${Q}<{Q}_c$, we observe distinct behaviors of the exponent of chaos for the charged particle around different sized BHs. The exponent for the case of small BHs initially increases slightly and then decreases as the temperature ${T}$ rises, while the exponent for that of large BHs decreases monotonically with increasing ${T}$. In contrast, for case of intermediate BHs, the exponent increases with the temperature. However, when ${Q}>{Q}_c$, the exponent is single valued. Moreover, there exists a terminal temperature at which the exponent approaches zero, signifying the disappearance of the unstable orbit. 

Figure \ref{f4-b} illustrates the scenario for ${Q}=0.12$. When ${T}_1<{T}<{T}_{p2}$, the exponent is a multivalued function of ${T}$, indicating the coexistence of large, intermediate and samll BH phases within this temperature range. Concurrently, in the regime ${T}_{p2}<{T}<{T}_{2}$, the exponent also shows two distinct branches corresponding to the intermediate and small BH phases. At ${T}={T}_{p2}$, the unstable orbit vanishes as the exponent approaches zero. Furthermore, Figure \ref{f4-d} demonstrates the functional relationship between the exponent and the horizon radius. Through a comparative analysis of Figure \ref{f4-b} and Figure \ref{f4-d}, we find the following: For small BHs, the exponent is significantly affected by the BH's charge, decreasing with increasing the charge. For large BHs, the influence of the charge becomes negligible. All curves in Figure \ref{f4-d} approach zero at nearly identical critical radii, regardless of the charge's values. This critical radius corresponds to the terminal temperature observed in Figure \ref{f4-b}. Therefore, the exponent for the charged particle also reveals the phase transition of the NLED BH.

\begin{figure}[h]
	\centering
	\includegraphics[width=9cm,height=8cm]{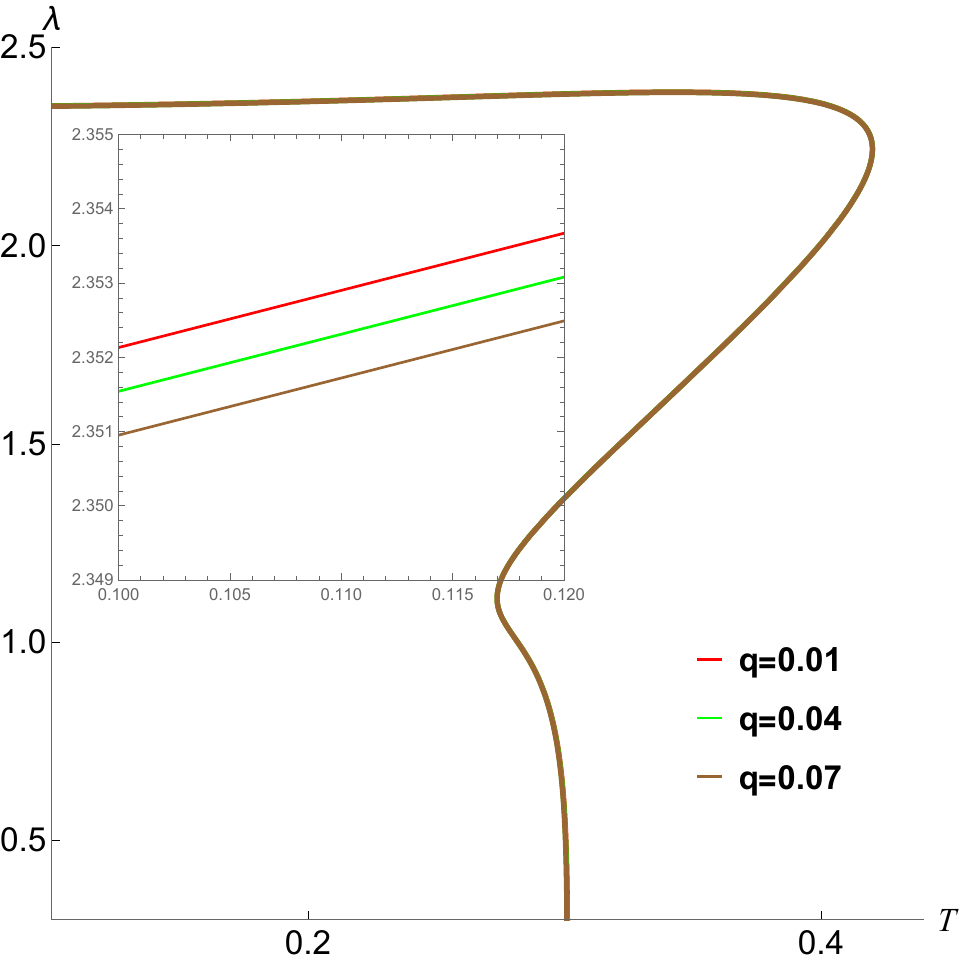}
	\caption{The variation of the LE of chaos for the different charged particle with respect to the temperature of the NLED BH.}
	\label{f5}
\end{figure}

The exponent of chaos for charged particles is also affected by the particle's charge. As depicted in Figure \ref{f5}, we illustrate the relationship between the exponent and the temperature for different values of the particle's charge. For the sake of generality, we set ${q}=0.01,0.04,0.07$, which is less than the BH's charge. Upon analyzing Figure \ref{f5}, we observe two distinct effects of the charge on the system. On one hand, although the overall influence of the charge on the exponent appears to be relatively minor in a broad sense, there is still a subtle yet discernible trend. Specifically, the exponent exhibits a slight decrease as the charge increases, but this change is well constrained within a certain limit. On the other hand, the unstable orbital position of the particle undergoes variations with changes in the charge, which in turn leads to corresponding alterations in the exponent value. In conclusion, within the range where  ${q}< {Q}$, the exponent remains a valid and reliable probe for detecting the phase transition in the NLED BH.

\subsection{Crtical exponents}\label{sec3.4}
In this subsection, we calculate the critical exponents of the phase transition for the NLED black hole to investigate the critical behavior of $\Delta\lambda$. The relationship between the critical exponent $\delta$ and $\Delta\lambda$ is defined as.

\begin{eqnarray}
\Delta \lambda \sim |T - T_{C}|^{\,\delta},
\label{eq3.4.1}
\end{eqnarray}

\noindent where $\lambda$ represents the chaos LE of particles. The horizon radius near the critical point can be written as

\begin{eqnarray}
r_+ = r_i(1 + \epsilon).
\label{eq3.4.2}
\end{eqnarray}

\noindent Here, $r_i$ is the horizon radius at the critical point and  $\left |\epsilon \ll1\right|$.The Hawking temperature $T$ can be written as a function of $r_+$. $T(r_+)$ can be rewritten as

\begin{eqnarray}
T\left( r_{+}\right) = T_{i}(1+\xi).
\label{eq3.4.3}
\end{eqnarray}

\noindent Where $\left |\xi \ll1\right|$, $T_{i}$ is the critical temperature. Near the critical point, we perform a Taylor expansion of $T(r_+)$ and obtain the following equation

\begin{eqnarray}
T(r_{+}) = T(r_{i}) 
+ \left( \frac{\partial T}{\partial r_{+}} \right)_{c} 
(r_{+} - r_{i}) 
+ \frac{1}{2} \left( \frac{\partial^{2} T}{\partial r_{+}^{2}} \right)_{c} 
(r_{+} - r_{i})^{2} 
+ \mathcal{O}(r_{i}) .
\label{eq3.4.4}
\end{eqnarray}

\noindent In the above equation, the subscript ‘c’ denotes the value at the critical point. At the critical point, $\left( \frac{\partial T}{\partial r_{+}} \right)_{c}=0$,and thus the second term on the right-hand side of the equation vanishes. Furthermore, neglecting the higher-order terms $\mathcal{O}(r_{i}$ and using Eq. (\ref{eq3.4.2}) and Eq. (\ref{eq3.4.4}), we obtain

\begin{eqnarray}
\epsilon^{2} = \frac{1}{2} \frac{T_{i} \xi}{r_{i}^{2}}
\left( \frac{\partial^{2} T}{\partial r_{+}^{2}} \right)_{c}.
\label{eq3.4.5}
\end{eqnarray}

Similarly, performing a Taylor expansion of $\lambda(r_{+})$ in the vicinity of the critical point $r_{i}$.

\begin{eqnarray}
\lambda(r_{+}) = \lambda(r_{i}) 
+ \left( \frac{\partial \lambda}{\partial r_{+}} \right)_{c} 
(r_{+} - r_{i}) + \mathcal{O}(r_{i}) .
\label{eq3.4.6}
\end{eqnarray}

\noindent Neglecting all higher-order terms in the above equation and using Eqs. (\ref{eq3.4.3}) (\ref{eq3.4.5}) and (\ref{eq3.4.6}), we obtain

\begin{eqnarray}
\lambda(r_{+}) - \lambda(r_{i}) 
= \left( \frac{\partial \lambda}{\partial r_{+}} \right)_{c}
\left( \frac{1}{2} \frac{\partial^{2} T}{\partial r_{+}^{2}} \right)_{r_{+}=r_{i}}^{-\tfrac{1}{2}}
\left( T - T_{i} \right)^{\tfrac{1}{2}} .
\label{eq3.4.7}
\end{eqnarray}

\noindent The critical exponent $\delta$ is defined as $\Delta \lambda \sim |T - T_{C}|^{\,\delta}$. From the above equation, we observe that the critical exponent is $\delta=1/2$. In \cite{LP47}, $\Delta \lambda \equiv \lambda_{L} - \lambda_{S}$ is regarded as the order parameter, where $\lambda_{L}$ and $\lambda_{S}$ denote the Lyapunov exponents of the large/small black hole phases. We obtain a consistent result with theirs, namely that the critical exponent is $\delta=1/2$.

\section{Chaos bound}\label{sec4}

Recently, Maldacena, Shenker and Stanford put forward that there is a universal upper bound for the LE in thermal quantum systems with a large number of degrees of freedom \cite{MSS}, 

\begin{eqnarray}
\lambda \leq \frac{2\pi T}{\hbar},
\label{eq2.19}
\end{eqnarray}

\noindent where ${T}$ represents the temperature of the system. Furthermore, within the context of single-particle systems, it has been further discovered that the upper bound of the exponent does not exceed the surface gravity of	the BH \cite{HT}. However, recent studies have revealed that, for certain BHs within specific parameter ranges, the chaos bound is violated \cite{LP72,LP73,LP74,LP75,LP76,SB1,SB2,JP,SBW,JJ}. When combined with the findings presented in the preceding sections, our results suggest that the exponent can serve as a reliable indicator of the BH phase transition. Consequently, we hypothesize the existence of an intrinsic connection between the chaos bound and the thermodynamic stability of BH. Specifically, our analysis indicates that the chaos bound may be violated during the phase transition, implying a potential link between chaotic dynamics and thermodynamic behavior in this system.

\begin{figure}[h]
  \centering
  \begin{subfigure}[b]{0.48\textwidth}
    \includegraphics[width=7cm,height=6cm]{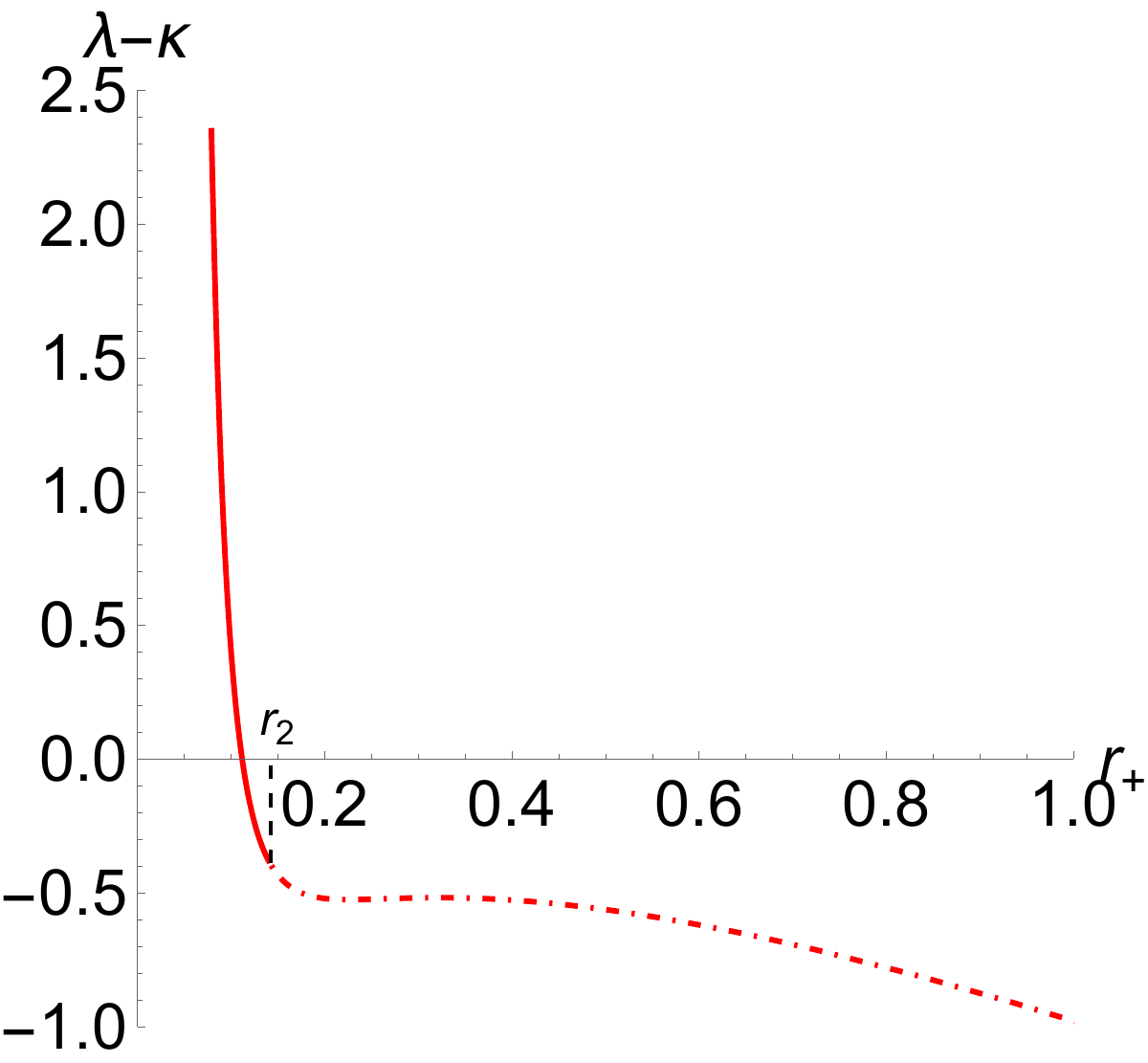}
    \caption{${Q}=0.12$}
    \label{f6-a}
  \end{subfigure}
  \hfill 
  \begin{subfigure}[b]{0.48\textwidth}
    \includegraphics[width=7cm,height=6cm]{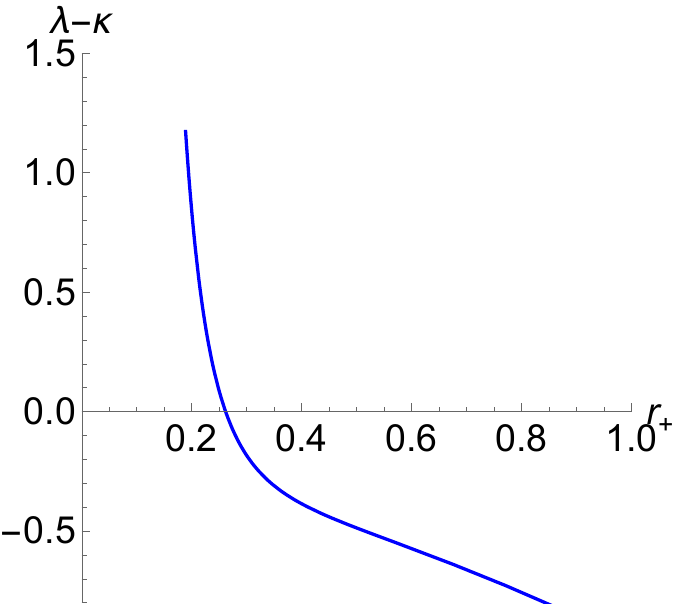}
    \caption{${Q}=0.20$}
    \label{f6-b}
  \end{subfigure}
  \caption{The variation of the LE of chaos for the massless particle with respect to the horizon radius of the NLED BH. In Figure \ref{f6-a}, the solid line corresponds to the case of the small BH branch, while the dashed line corresponds to the case of the non-small BH (intermediate and large BHs) branch.}
  \label{f6}
\end{figure}

\begin{figure}[H]
  \centering
  \begin{subfigure}[b]{0.48\textwidth}
    \includegraphics[width=7cm,height=6cm]{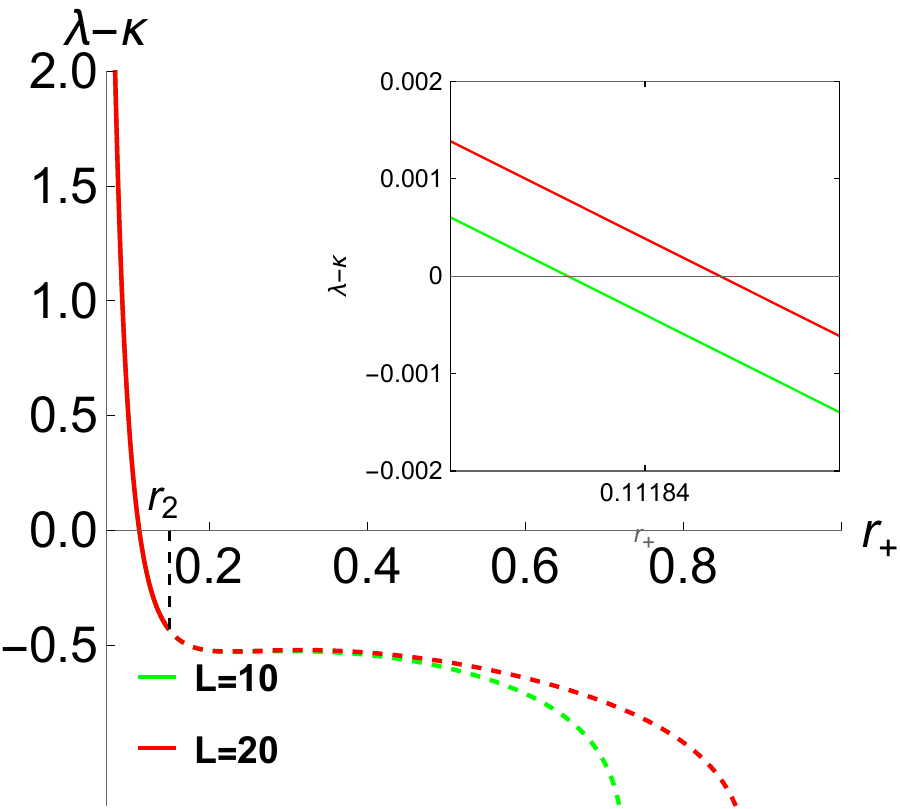}
    \caption{${Q}=0.12$}
    \label{f7-a}
  \end{subfigure}
  \hfill 
  \begin{subfigure}[b]{0.48\textwidth}
    \includegraphics[width=7cm,height=6cm]{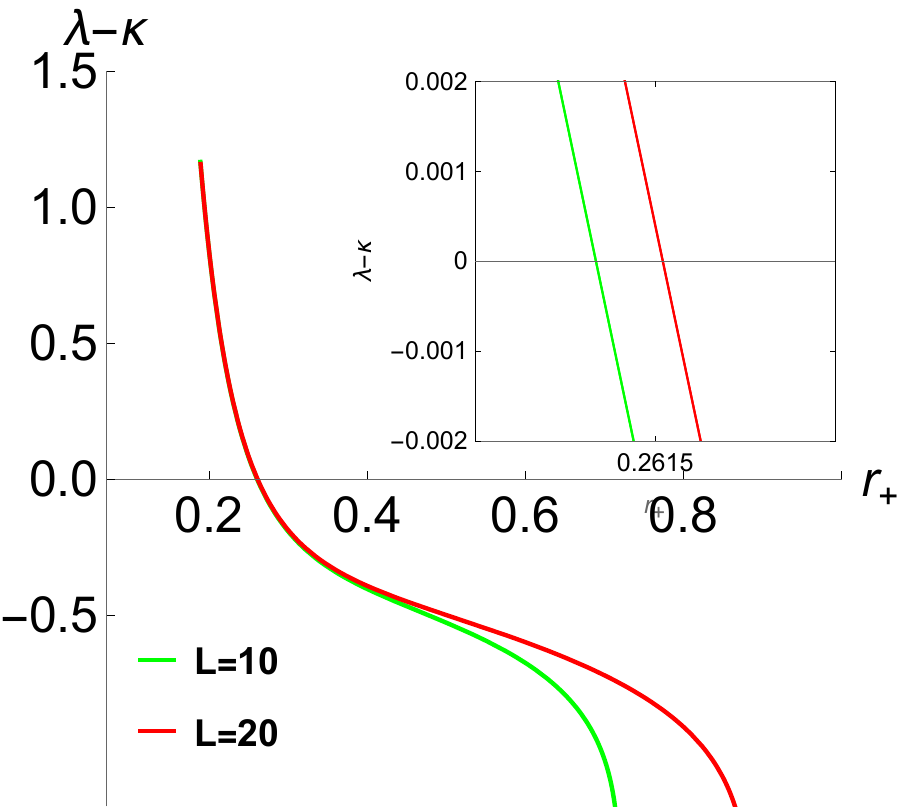}
    \caption{${Q}=0.20$}
    \label{f7-b}
  \end{subfigure}
  \caption{The variation of the LE of chaos for the charged particle with respect to the horizon radius of the NLED BH. In Figure \ref{f7-a}, the solid line corresponds to the case of the small BH branch, while the dashed line corresponds to the case of the non-small BH (intermediate and large BHs) branch.}
  \label{f7}
\end{figure}

In Figures \ref{f6} and \ref{f7}, we elucidate the relationship between the LE and the surface gravity ($\kappa$ normalized by $\frac{2\pi T}{\hbar}$) of the NLED BH, considering both the massless (Figure \ref{f6}) and charged (Figure \ref{f7}) particles, respectively.  It is important to emphasize that: ${r}_2$ in subfigures \ref{f6-a} and \ref{f7-a} denotes a critical radius at which the BH undergoes a phase transition from the small BH phase to the non-small BH phase, and this corresponds to the temperature $T_2$ depicted in Figure \ref{f4-b}. The left hand side of ${r}_{2}$ represents the stable branch of small BHs. $\lambda-\kappa>0$ indicate a violation of the chaos bound. Figure \ref{f6} elucidates the relationship between the LE characterized chaos for the massless particle and the horizon radius. As can be discerned from the figure, when the radius is less than ${r}_2$, a violation of the chaos bound occurs, indicating that the violation manifests in the spacetime of the stable small BH phase. Figure \ref{f6-b} illustrates the scenario where $Q=0.2>Q_c$. In this case, the NLED BH does not undergo a phase transition, yet we still observe a violation of the chaos bound. Moreover, by comparing it with Figure \ref{f6-a}, we find that an increase in the BH's charge enlarges the threshold radius corresponding to the occurrence of the violation.

Figure \ref{f7} reveals the relationship between the LE characterized chaos for the charged particle and the horizon radius. In Figure \ref{f7-a}, as the particle's angular momentum increases, the violation region of the chaos bound also expands. It is noteworthy that, similar to Figure \ref{f6-a}, the violation occurs solely in the spacetime of the stable small BH phase, and the violation region always remains below the critical value $r_2$ at which the small BH phase transits to other phases. Figure \ref{f7-b} presents the situation where the BH does not experience a phase transition. We discover that an increase in the angular momentum also enlarges the region of the violation. Furthermore, by comparing it with Figure \ref{f7-a}, we find that an increase in the BH's charge increases the threshold radius corresponding to the violation. 

Figures \ref{f6} and \ref{f7} collectively demonstrate that the violation of the chaos bound occurs regardless of whether the BH undergoes a phase transition. However, it is important to note that such a violation occurs only in the spacetime of small BHs with a thermodynamically stable phase.

\section{Conclusions}

In this paper, we studied the LEs of chaotic motion for both massless and charged particles orbiting around the NLED BH, and discussed their connections with the phase transition and the chaos bound of this BH. The findings demonstrate that these exponents can effectively detect the phase transition. 

In \cite{LP78}, the authors have discovered that the violation of the chaos bound occurs within the stable phase of the BH. Our study further reveals that when a phase transition takes place, the violation occurs in the stable branch of the small BH. Moreover, regardless of whether a phase transition occurs, the phenomenon of the violation manifest itself. Simultaneously, during the phase transition process, there exists a critical value for the horizon. Only when the horizon is smaller than this critical value the violation of the chaos bound occur. Additionally, an increase in the angular momentum of the charged particle expands the region where the chaos bound is violated.

\appendix
\section {Lyapunov exponent of massless particles}

When a test particle moves in a circle around the equator of the black hole, the Lagrangian is 
\begin{eqnarray}
\mathcal{L} = \frac{1}{2}\left(-F\dot{t}^2+\frac{\dot{r}^2}{F} +r^2\dot{\phi}^2\right),
\label{eqA.1}
\end{eqnarray}

\noindent Using $ \pi_{\mu}=\frac{\partial\mathcal{L}}{\partial\dot{x}}$, we get the generalized momentum as follows:
\begin{eqnarray}
\pi_t = \frac{\partial\mathcal{L}}{\partial\dot{t}} = -F\dot{t}=-E, \quad\quad \pi_r = \frac{\partial\mathcal{L}}{\partial\dot{r}} = \frac{\dot{r}}{F}, \quad\quad \pi_{\phi} =\frac{\partial\mathcal{L}}{\partial\dot{\phi}} = r^2 \dot{\phi}=L.
\label{eqA.2}
\end{eqnarray}

\noindent There $E$ and $L$ denote the energy and angular momentum of the particle. Thus the Hamiltonian of the particle is
\begin{eqnarray}
H = \frac{1}{2F}({-\pi_{t}^2+\pi_r^2F^2+ \pi^2_{\phi}r^{-2}F}).
\label{eqA.3}
\end{eqnarray}

\noindent From the Hamiltonian, the motion equations of the particle are gotten, which are:
\begin{eqnarray}
\dot{t} &=& \frac{\partial H}{\partial \pi_t}=-\frac{\pi_t}{F}, \quad  \dot{\pi_t}= -\frac{\partial H}{\partial t} =0 ,
\quad \dot{r} = \frac{\partial H}{\partial \pi_r}= \pi_r F, \nonumber\\
\dot{\pi_r} &=& -\frac{\partial H}{\partial r} =-\frac{1}{2}\left[\pi^2_r F^{\prime} +\frac{\pi_t^2F^{\prime}}{F^2} -\pi^2_{\phi}r^{-4}({r^2})^{\prime}\right], \nonumber\\
\dot{\phi} &=& \frac{\partial H}{\partial \pi_{\phi}}= \frac{\pi_{\phi}}{r^2}, \quad  \dot{\pi_\phi}= -\frac{\partial H}{\partial \phi} =0.
\label{eqA.4}
\end{eqnarray}

\noindent Here, we again apply the normalization condition of the particle's four-velocity, where $\eta =0$ corresponds to the case of massless particles, yielding the constraint condition.
\begin{eqnarray}
\pi_t=-\sqrt{F( \pi_r^2N + \pi_{\phi}^2r^{-2})}.
\label{eqA.5}
\end{eqnarray}

\noindent Similarly, we derive the relationship between the radial coordinate and time for the massless particle.
\begin{eqnarray}
\frac{dr}{dt} &=& \frac{\dot{r}}{\dot{t}} =-\frac{\pi_rF^2}{\pi_t}, \nonumber\\
\frac{d\pi_r}{dt} &=& \frac{\dot{\pi_r}}{\dot{t}} =\frac{1}{2}\left[\frac{\pi^2_r FF^{\prime}}{\pi_t}+\frac{\pi_t F^{\prime}}{F} -\frac{\pi^2_{\phi}r^{-4}({r^2})^{\prime}F}{\pi_t}\right].
\label{eqA.6}
\end{eqnarray}

\noindent For massless particle, Eq. (\ref{eqA.6}) is reformulated by using constraint Eq. (\ref{eqA.5}) and becomes 
\begin{eqnarray}
F_1 &=& \frac{\pi_rF^2}{\sqrt{F(\pi_r^2F + \pi_{\phi}^2r^{-2})}}, \nonumber\\
F_2 &=& -\frac{\pi^2_r (F^2)^{\prime}}{2\sqrt{F(\pi_r^2F + \pi_{\phi}^2r^{-2})}} -\frac{\pi^2_{\phi}(r^{-2}F)^{\prime}}{2\sqrt{F(\pi_r^2F + \pi_{\phi}^2r^{-2})}}.
\label{eqA.7}
\end{eqnarray}

Considering the circular motion of the particle, its constraint condition satisfies:
\begin{eqnarray}
\pi_r=\frac{d\pi_r}{dt}=0.
\label{eqA.8}
\end{eqnarray}

\noindent For a massless particle, when $\eta =0$, it naturally reduces to the photon sphere case, and we determine its orbit by $F_2=0$. Using Eq. (\ref{eq2.15}) and (\ref{eqA.7}), derive the Jacobian matrix in phase space $(r, \pi_r)$ and compute its eigenvalues to obtain the Lyapunov exponent. The elements of the Jacobian matrix are as follows 
\begin{eqnarray}
K_{11} &=& \frac{\partial F_1}{\partial r}=0, \nonumber\\
K_{12}&=& \frac{\partial F_1}{\partial \pi_r} =\frac{F^2}{\sqrt{F(\pi_{\phi}^2r^{-2})}},\nonumber\\
K_{21}&=& \frac{\partial F_2}{\partial _r} = 
-\frac{\pi^2_{\phi}(r^{-2}F)^{\prime\prime}}{2\sqrt{F(\pi_{\phi}^2r^{-2})}} + \frac{\left[\pi_{\phi}^2(r^{-2}F)^{\prime}\right]^2}{4\left[F(\pi_{\phi}^2r^{-2})\right]^{3/2}},\nonumber\\
K_{22} &=& \frac{\partial F_2}{\partial \pi_r}=0.
\label{eqA.9}
\end{eqnarray}

\noindent The chaos exponent of the massless particle can be obtained by calculating the eigenvalues of the matrix, which is:
\begin{eqnarray}
\lambda^2 = -\frac{1}{2}\frac{F(r^{-2}F)^{\prime\prime}}{r^{-2}}+ \frac{1}{4}\frac{F\left[(r^{-2}F)^{\prime}\right]^2}{F(r^{-2})^2}.
\label{eqA.10}
\end{eqnarray}

\end{document}